\documentstyle[11pt]{article}

\def\Bagd{B_\al\,^{\ga\de}}
\def\eth{e_\al\,^\vartheta}
\def\Habg{H_{\al\be}\,^\ga}

\def\Rabgd{R_{\al\be}\,^{\ga\de}}
\def\Rab{R_{\al\be}}
\def\Tabg{T_{\al\be}\,^\ga}

\def\bib{\bibitem}

\def\dem{\det e^{-1}}

\def\intf{\int d^{4}x\,}

\def\lar{\longrightarrow}
\def\lbr{\lbrack}

\def\pa{\partial}

\def\rbr{\rbrack}

\def\al{\alpha}
\def\be{\beta}
\def\ga{\gamma}
\def\Ga{\it\Gamma}
\def\de{\delta}
\def\ep{\varepsilon}
\def\ze{\zeta}
\def\th{\vartheta}
\def\ka{\kappa\,}

\def\La{{\it\Lambda}}

\def\Si{\Sigma}
\def\va{\varphi}
\def\om{\omega}

\def\beq{\begin{equation}}
\def\eeq{\end{equation}}
\def\bed{\begin{displaymath}}
\def\eed{\end{displaymath}}
\def\beqq{\begin{eqnarray}}
\def\eeqq{\end{eqnarray}}
\def\bedd{\begin{eqnarray*}}
\def\eedd{\end{eqnarray*}}

\title{\huge{\bf{Schwarzschild Solution within a new renormalizable SO(1,3) Gauge Field Theory of Gravitation}}}

\author{\vspace{0.5cm}\\
C. Wiesendanger\\
Aurorastr. 24\\
8032 Zurich, Switzerland\\
{\it chwiesendanger@bluewin.ch}}
\date{May 5th, 2024}

\begin{document}

\maketitle

\begin{abstract}
The existence of the Schwarzschild solution is demonstrated within a new {\bf SO(1,3)\/} gauge field theory whose fundamental gauge field $B$ is of dimension one allowing for the renormalizability of the full quantum theory. On the other hand the classically dominant term of the gauge field action can be expressed in terms of a naturally emerging Vierbein of dimension zero. In fact this action term is the same functional of the emerging Vierbein $e[B]$ as is the Einstein-Hilbert action of the Vierbein in General Relativity (GR), albeit with $e[B]$ subject to a constraint. To solve the Einstein equations a spherically symmetric Ansatz for $e[B]$ respecting the constraint is determined and transformed into the standard metric used to calculate the Schwarzschild solution. This allows to prove the latter's existence making the present theory a perfectly viable theory of gravitation equivalent to GR at the classical level.\\
\end{abstract}

\clearpage

\section{Introduction}

\paragraph{}
In a series of papers \cite{chw1,chw2,chw3} we have developed a gauge field theory of the Lorentz group {\bf SO(1,3)\/} describing gravity potentially at the classical and the quantum level.

There are some key differences of the present theory to gauge theories of gravity discussed in the literature which come with promising features missing in the usual approaches \cite{hehl1,hehl2,cho1,cho2,uti,kib}. 

In those either the translation group is gauged leading to a non-renormali-zable theory of gravitation which cannot integrate spin easily, or both the translation and the Lorentz groups are gauged with torsion and curvature appearing and automatically integrating spin, but still not renormalizable due to the dimension-zero fields related to the translation group. In all these approaches it is the translation group which accounts for the spacetime-related local transformations whereas the local Lorentz group only acts on spin degrees of freedom.

The construction of the present theory follows the same philosophy and technical steps as the construction of gauge theories used in particle physics albeit with the complication of a gauge group acting on representation spaces of infinite dimension. In fact the Lorentz group and the respective gauge fields act equally on spacetime and spin degrees of freedom of all other fields and are of dimension one which as a consequence points to renormalizability \cite{chw1}. In fact the generating functional for Green functions built from the most general combination of dimension-zero, -two and -four Lagrangians in the gauge fields has been shown to be perturbatively renormalizable \cite{chw3}.

There are obviously many open points to be still addressed amongst which are the existence of a Schwarzschild-type solution or more generally the demonstration of the theory's equivalence to General Relativity (GR) at the classical level to make it a viable theory of gravitation. Or at the quantum level -- beyond renormalizability -- the question of unitarity given that {\bf SO(1,3)\/} is a non-compact group which makes the construction of a proper Hilbert space of field quanta a challenge \cite{chw2}.

In this paper we address the point whether the present Lorentz gauge field theory is a viable theory of gravitation matching key results of GR, and in particular whether the Schwarzschild solution exists within this theory.

To get there we revisit the essentials of the new {\bf SO(1,3)\/} gauge field theory in the first four sections of this paper and specifically link the dominant dimension-two part of the action to the Einstein-Hilbert action, hence establishing the connection to GR.

Having connected the dominant term in the action in the classical realm to GR, however, is not sufficient to automatically assure the existence of a Schwarzschild-type solution because of what emerges as a Vierbein-type of field in the process is subject to a specific constraint. So in section six we establish a general spherically symmetric Vierbein Ansatz for a solution to the field equations respecting this constraint and in section seven we calculate the corresponding metric.

Finally in section eight taking this metric and the corresponding line element we demonstrate the existence of the Schwarzschild solution following well-known steps in the literature.

\section{{\bf SO(1,3)\/} Gauge Transformations and the Covariant Derivative}

\paragraph{}
In this and the next three sections we recapitulate the essentials of the {\bf SO(1,3)\/} gauge field theory developed in \cite{chw1} starting with {\bf SO(1,3)\/} gauge transformations and the covariant derivative necessary to build further covariant objects.

First we introduce the relevant notations for both the local Lorentz gauge group {\bf SO(1,3)\/} and algebra {\bf so(1,3)\/} at the core of the theory. For our purpose it is sufficient to work with infinitesimal group elements ${\bf 1} + \Theta_\om (x)$ where
\beq \Theta_\om (x) = \frac{i}{2}\om^{\ga\de} (x) (L_{\ga\de} + \Si_{\ga\de}) \in {\bf so(1,3)\/}
\eeq
is a local element of the gauge algebra {\bf so(1,3)\/}. Here, $x$ denotes a point in Minkowski spacetime {\bf M\/}$^{4}$ $\equiv$ ({\bf R\/}$^{4}$,$\eta$) and $\eta=\mbox{diag}(-1,1,1,1)$ the flat spacetime metric with which indices are raised and lowered throughout this paper. Indices $\al,\be,\ga,...$ denote quantities defined on {\bf M\/}$^{4}$ which transform covariantly under global Lorentz transformations.

The six infinitesimal gauge parameters parametrizing the Lorentz algebra are given by 
$\om^{\ga\de} (x) = - \om^{\de\ga} (x)$. The spacetime-related algebra generators acting on field space are given by
\beq L_{\ga\de} = -L_{\de\ga} = -i(x_{\ga}\pa_{\de}-x_{\de}\pa_{\ga})
\eeq
whereby Lorentz transformations are treated as "inner" transformations \cite{chw1}.

Finally $\Si_{\ga\de}$ denote the generators of an arbitrary finite-dimensional spin representation of the Lorentz algebra.

Generators $J_{\al\be}$ of any representation obey the commutation relations
\beq \lbr J_{\al\be},J_{\ga\de}\rbr
= i\{\eta_{\al\ga}J_{\be\de}-\eta_{\be\ga}J_{\al\de}
+\eta_{\be\de}J_{\al\ga}-\eta_{\al\de}J_{\be\ga}\}
\eeq
which in effect define the Lorentz algebra {\bf so(1,3)\/}.

Next we turn to the action of a local infinitesimal group element ${\bf 1} + \Theta_\om (x)$ on a generic field $\va(x)$ living in an arbitrary finite-dimensional spin representation of the Lorentz group. 

The variations of the spacetime coordinates and the field under "inner" gauge transformations are given by
\beqq x^\al \lar x'^\al&=&x^\al \\
\va(x)\lar \va'(x)&=&\va(x) + \de_\om \va(x) \nonumber
\eeqq
with
\beqq & & \de_\om \va (x) = (\Theta_\om (x) \va)(x) \\
&=& -\om^{\ga\de}(x)  x_\de \pa_\ga \va (x) 
+\frac{i}{2}\om^{\ga\de}(x)  \Si_{\ga\de} \va (x). \nonumber
\eeqq
Hereby "inner" means that spacetime coordinates do not transform under a gauge transformation and that such a transformation and its generators act on field degrees of freedom only, be them finite-dimensional ones labeled by some discrete spinor indices or infinite-dimensional ones labeled by the continuous spacetime "index" $x$.

Finally we need to introduce a covariant derivative $\nabla_\al(x)$ in order to define locally gauge-covariant expressions. This derivative is defined by
\beq \Big(\nabla_\al(x) \va \Big)' = \nabla'_\al(x) \va'
\eeq
with a prime referring to a gauge-transformed quantity.

Taking the usual Ansatz we decompose $\nabla_\al(x)$ into $\pa_\al$ and a gauge field $B_\al(x)$
\beqq \nabla^B_\al &=& \pa_\al + B_\al(x) \nonumber \\
B_\al &=& \frac{i}{2}\Bagd(x) (L_{\ga\de} + \Si_{\ga\de}) \\
&=& -\Bagd(x) x_\de \pa_\ga
+\frac{i}{2}\Bagd(x) \Si_{\ga\de} \nonumber
\eeqq
living in the Lorentz algebra {\bf so(1,3)\/} with $\Bagd$ antisymmetric in the inner indices $\ga$ and $\de$.

To further develop the theory we rewrite the covariant derivative $\nabla^B_\al$
\beqq \nabla^B_\al &=& \left( \eta_\al\,^\ga - \Bagd x_\de \right) \pa_\ga
+\frac{i}{2}\Bagd \Si_{\ga\de} \\
&=& d^B_\al + {\bar B}_\al \nonumber
\eeqq
thereby introducing the expression
\beq  \label{9}
e_\al\,^\th[B] \equiv \eta_\al\,^\th - B_\al\,^{\th\ze}x_\ze
\eeq
to keep the algebraic expressions simple and adding the short-hand notations
\beq d^B_\al \equiv e_\al\,^\th[B]\, \pa_\th,
\quad {\bar B}_\al \equiv \frac{i}{2}\Bagd \Si_{\ga\de}.
\eeq

The $e_\al\,^\th[B]$ above resembles a Vierbein. But by its definition Eqn.(\ref{9}) it is a functional of the fundamental dynamical gauge field $\Bagd$ in our theory.

In \cite{chw1} we have elaborated in depth why $e_\al\,^\th[B]$ not being a fundamental field is so crucial for the further development of the theory which turns out to be both equivalent to General Relativity in the classical realm, and to be renormalizable after quantization \cite{chw3}.

Furthermore
\beq  \label{11}
e_\al\,^\th[B]\, x_\th = x_\al - B_\al\,^{\th\ze} x_\ze x_\th = x_\al
\eeq
by its very definition in Eqn.(\ref{9}) acts on spacetime vectors $x_\al$ as a projector due to the antisymmetry of $B_\al\,^{\th\ze}$ in the inner indices $\th$ and $\ze$.

We note that the condition Eqn.(\ref{11}) is covariant under local {\bf SO(1,3)\/} gauge transformations. This is easily demonstrated given the explicit form of local {\bf SO(1,3)\/} transformations acting on the gauge fields $\Bagd$ as
\beq
\de_\om\Bagd= -\om^{\eta\ze}x_\ze \pa_\eta \Bagd - d^B_\al \om^{\ga\de} + \om_\al\,^\be B_\be\,^{\ga\de} + \om^\ga\,_\eta B_\al\,^{\eta\de} + \om^\de\,_\eta B_\al\,^{\ga\eta} \eeq
and acting on $\eth [B]$ as
\beq
\de_\om\eth [B] = -\om^{\eta\ze}x_\ze \pa_\eta \eth [B]
+ e_\al\,^\ep [B] \pa_\ep (\om^{\th\ze}x_\ze) 
+\om_\al\,^\be e_\be\,^\th [B] \eeq
with the spacetime coordinates $x_\al$ left invariant, i.e. $\de_\om x_\al = 0$.

\section{{\bf SO(1,3)\/} Field Strength Tensor and its Components}

\paragraph{}
In this section we revisit the construction of the field strength tensor and its components.

We first turn to defining the field strength $G$ operator acting on field space
\beq G_{\al\be} [B]\equiv \left[ \nabla^B_\al,\nabla^B_\be \right]
\eeq
and to expressing it in terms of the gauge field $B$
\beqq G_{\al\be} [B]&=& \left[\, d^B_\al ,\, d^B_\be\, \right] 
+ d^B_\al {\bar B}_\be - d^B_\be {\bar B}_\al \\
&+& \left[{\bar B}_\al , {\bar B}_\be \right] + ( B_{\al\be}\,^\eta - B_{\be\al}\,^\eta ) \nabla^B_\eta \nonumber
\eeqq
where the last term accounts for the vector nature of $\nabla^B_\al$.

Next we calculate
\beq \left[\, d^B_\al ,\, d^B_\be\, \right] = \left( e_\al\,^\ze [B]\,\pa_\ze e_\be\,^\eta [B]-
e_\be\,^\ze [B]\,\pa_\ze e_\al\,^\eta [B] \right)\pa_\eta = \Habg [B]\, d^B_\ga
\eeq
defining
\beq \Habg [B] \equiv e^\ga\,_\eta [B] \left( e_\al\,^\ze [B]\,\pa_\ze e_\be\,^\eta [B]-
e_\be\,^\ze [B]\,\pa_\ze e_\al\,^\eta [B] \right)
\eeq
where we have made use of the inverse $e^\ga\,_\eta [B]$ with
$e^\ga\,_\eta [B]\, e_\ga\,^\ze [B]=\de_\eta\,^\ze$ assuming that $e_\al\,^\ze [B]$ is non-singular, i.e. $\det e[B]\neq 0$.

Now we can express
\beqq G_{\al\be} [B] &=& \Big(\Habg [B]+ B_{\al\be}\,^\ga 
- B_{\be\al}\,^\ga \Big) \nabla^B_\ga  \nonumber \\
&+& d^B_\al {\bar B}_\be - d^B_\be {\bar B}_\al + [{\bar B}_\al ,{\bar B}_\be] - \Habg [B]\, {\bar B}_\ga \\
&=& - \Tabg [B]\, \nabla^B_\ga + \Rab [B]  \nonumber 
\eeqq
in terms of the covariant quantities
\beq \Tabg [B] \equiv -(B_{\al\be}\,^\ga - B_{\be\al}\,^\ga) - \Habg [B]
\eeq
and
\beqq \Rab [B] &\equiv& \frac{i}{2} \Rabgd [B] \, \Si_{\ga\de} \nonumber \\
\Rabgd [B] &=& d^B_\al  B_\be\,^{\ga\de} - d^B_\be B_\al\,^{\ga\de}
+\, B_\al\,^{\ga\eta}\, B_{\be\eta}\,^\de \\ 
&-& B_\be\,^{\ga\eta}\, B_{\al\eta}\,^\de 
- H_{\al\be}\,^\eta [B]\, B_\eta\,^{\ga\de}. \nonumber
\eeqq
The transformation behaviour of the various quantities under local Lorentz gauge transformations has been given in \cite{chw1}.

We finally point out the geometrical significance of all, the gauge field $B$, field strength $G$ and its components $R$ and $T$. It has been clarified in terms of a Banach fibre bundle structure with trivial base manifold {\bf M\/}$^{4}$ and infinite-dimensional fibres for the various fields in \cite{chw1}. More specifically $R$ corresponds to the Riemann curvature tensor and $T$ to the torsion tensor in differential geometrical terms.

\section{Most General Renormalizable {\bf SO(1,3)\/} Gauge Field Action}

\paragraph{}
In this section we revisit the most general renormalizable gauge-invariant action for the gauge field $B$.

Let us turn to writing down the most general renormalizable gauge-invariant action for the gauge field $B$ the full renormalizabilty of which has been proven in \cite{chw3}. It contains dimension-zero, -two and -four contributions $S^{(0)}_G [B]$, $S^{(2)}_G [B]$ and $S^{(4)}_G [B]$ respectively. Without giving further details we display the different terms determined in \cite{chw1} starting with
\beq S^{(0)}_G [B] = \La\, \intf \dem [B] \eeq
where $ \La $ is a constant of dimension $[\La] = 4$ which can be interpreted as cosmological constant.

The most general dimension-two contribution reads
\beqq \label{22}
S^{(2)}_G [B] &=& \frac{1}{\ka}\, \intf \dem [B]
\Big\{\al_1\, R_{\al\be}\,^{\al\be} [B] \nonumber \\
&+& \al_2\, T_{\al\be\ga} [B]\, T^{\al\be\ga} [B]
+ \al_3\, T_{\al\be\ga} [B]\, T^{\ga\be\al} [B] \\
&+&  \al_4\, T_\al \,^{\ga\al} [B]\, T_{\be\ga} \,^\be [B]
+  \al_5\, \nabla^B_\al T_\be \,^{\al\be} [B] \Big\}. \nonumber
\eeqq
$\frac{1}{\ka} = \frac{1}{16\pi \it\Ga}$ has mass-dimension $[ \frac{1}{\ka} ] = 2$ with ${\it\Ga} $ denoting the Newtonian gravitational constant. In fact $\ka$ is linked to the square of the Planck mass by $\frac{16\pi\hbar}{\ka} = \frac{\hbar}{\it\Ga} = m_P^2$. The $ \al_i $ above are constants of dimension $[\al_i] = 0$.

Finally, the most general dimension-four contribution reads
\beqq S^{(4)}_G [B] &=& \intf \dem [B]
\Big\{ \be_1\, R_{\al\be}\,^{\ga\de} [B]\, R^{\al\be}\,_{\ga\de} [B] \nonumber \\
&+&  \be_2\, R_{\al\ga}\,^{\al\de} [B]\, R^{\be\ga}\,_{\be\de} [B] 
+ \be_3\, R_{\al\be}\,^{\al\be} [B] \,R_{\ga\de}\,^{\ga\de} [B] \nonumber \\
&+& \be_4\, \nabla_B^\ga \nabla^B_\de R_{\al\ga}\,^{\al\de} [B]
+ \be_5\, \nabla_B^\ga \nabla^B_\ga R_{\al\be}\,^{\al\be} [B] \nonumber \\
&+& \dots \nonumber \\
&+& \ga_1\, \nabla^B_\ga T_{\al\be\de} [B]\,
\nabla_B^\ga T^{\al\be\de} [B]
+ \ga_2\, \nabla^B_\ga T_{\al\be\de} [B]\, 
\nabla_B^\ga T^{\de\be\al} [B] \\
&+& \dots \nonumber \\
&+& \ga_j\, T^4-\mbox{terms} \nonumber \\
&+& \dots \nonumber \\
&+& \de_k\, R\,T^2-\mbox{terms},\, R\,\nabla^B\, T-\mbox{terms} \nonumber \\
&+& \dots \Big\} \nonumber 
\eeqq 
with $ \be_i $, $ \ga_j $, $ \de_k $ constants of dimension $[\be_i] = [\ga_j] = [\de_k] = 0$. We note that it is $S^{(4)}_G [B]$ which makes the theory renormalizable.

\beq \label{24} S_G [B] = S^{(0)}_G [B] + S^{(2)}_G [B] + S^{(4)}_G [B]
\eeq
is the most general action of dimension $\leq 4$ in the gauge fields $\Bagd$ and their first  derivatives $\pa_\be \Bagd$ which is locally Lorentz invariant and by inspection renormalizable by power-counting.

The actual proof of renormalizability is given in \cite{chw3}. It requires the much more involved demonstration that counterterms needed to absorb infinite contributions to the perturbative expansion of the effective action of the full quantum theory are again of the form Eqn.(\ref{24}) with possibly renormalized constants.

\section{Gauge Field Action Dominant in the Classical Realm and its Equivalence to the Einstein-Hilbert Action}

\paragraph{}
In this section focusing on the classical realm we zoom in on the dominant term of the gauge field action and revisit its equivalence to the vacuum Einstein-Hilbert action in GR.

Inspection of the constants in $S_G [B]$ from Eqn.(\ref{24}) above indicates that the term $S^{(2)}_G [B]$ being of order Planck mass squared dominates the action for energies much smaller than the Planck mass.

We furthermore note that for the choice of constants in $S^{(2)}_G [B]$ in Eqn.(\ref{22}) above
\beq \al_1 = 1,\quad \al_2 = -\frac{1}{4},\quad \al_3 = -\frac{1}{2},\quad
\al_4 = -1,\quad \al_5 = 2
\eeq
$S^{(2)}_G [B]$ is mathematically equivalent to the Einstein-Hilbert action in the Vierbein approach \cite{chw1}.

To demonstrate this we have introduced in \cite{chw1} a new field $C_\al\,^{\be\ga}$ 
associated to the original gauge field $\Bagd$ by demanding that
\beq \Tabg [C] \equiv -\left(C_{\al\be}\,^\ga - C_{\be\al}\,^\ga\right) - \Habg [B] =^{\!\!\!\! !}\,\, 0. \eeq
We note that $C_\al\,^{\be\ga} [B]$ corresponds to the torsion-free Levi-Civita connection in GR .

Solving for $C_\al\,^{\be\ga} [B]$ we find \cite{chw1}
\beqq
\!\!\!\!\!\!\!\!\!C_\al\,^{\be\ga} [B] =  
\! \!\!&-&  \! \!\!\frac{1}{2} 
\left( H_\al\,^{\be\ga} [B] - H_\al\,^{\ga\be} [B] - H^{\be\ga}\,_\al [B] \right)
\nonumber \\
= \! \!\!&-&  \! \!\!\frac{1}{2}\, e^\ga\,_\eta [B] \left( e_\al\,^\ze [B]\,\pa_\ze e^{\be\eta} [B]-
e^{\be\ze} [B]\,\pa_\ze e_\al\,^\eta [B] \right) \\
  \! \!\!&-&  \! \!\! \frac{1}{2}\, e^\be\,_\eta [B] \left( e_\al\,^\ze [B]\,\pa_\ze e^{\ga\eta} [B]-
e^{\ga\ze} [B]\,\pa_\ze e_\al\,^\eta [B] \right) \nonumber \\
  \! \!\!&-&  \! \!\! \frac{1}{2}\, e_{\al\eta} [B] \left( e^{\be\ze} [B]\,\pa_\ze e^{\ga\eta} [B]-
e^{\ga\ze} [B]\,\pa_\ze e^{\be\eta} [B] \right) \nonumber
\eeqq
which makes $C_\al\,^{\be\ga} [B]$ functionally dependent on $B_\al\,^{\be\ga}$ through $e_\al\,^\th [B]$ only.

Note that $\Bagd$ can be written as a very convoluted functional of itself as shown in \cite{chw1}
\beq
\Bagd = C_\al\,^{\ga\de} \Big[e [B]\Big] - \frac{1}{2} \left( T_\al\,^{\ga\de} [B]
- T_\al\,^{\de\ga} [B] - T^{\ga\de}\,_\al [B] \right).
\eeq

For $S^{(2)}_G \bigg[ C\Big[e[B]\Big] \bigg]$ with $\! 1 \! = \! \al_1 \! = \! -4 \, \al_2 \! = \! -2 \, \al_3 $ $ \! = \! -\al_4 \! = \! \frac{1}{2} \, \al_5$  we have demonstrated in \cite{chw1} that
\beqq \label{29}
S^{(2)}_G \bigg[ C\Big[e[B]\Big] \bigg] &=&\frac{1}{\ka}\, \intf \dem [B]\, R^{\al\be}\,_{\al\be} \bigg[ C\Big[e[B]\Big] \bigg] \\
&=& S_{EH} [e] \nonumber
\eeqq
which -- in spite of fundamentally being a functional of $\Bagd$ and $\pa_\be \Bagd$ --  depends functionally on $\eth [B]$ and its first derivatives only and is the same functional of $e_\al\,^\th$ as is the Einstein-Hilbert action in the Vierbein approach to GR.

This means that $S^{(2)}_G \bigg[ C\Big[e[B]\Big] \bigg]$ not only is invariant under local {\bf SO(1,3)\/} gauge transformations, but also under diffeomorphisms which allows us to deploy the full mathematical apparatus of GR in both the Vierbein and metric approaches to demonstrate the existence of the Schwarzschild solution in our theory.

Finally we determine the vacuum field equations by varying Eqn.(\ref{29})
\beq
\frac{\de S^{(2)}_G}{\de \Bagd} = \int \dem [B]
\frac{\de S^{(2)}_G}{\de \eth [B]}\, \frac{\de \eth [B]}{\de \Bagd} = 0
\eeq
which are equivalent to
\beq
\frac{\de S^{(2)}_G}{\de \eth [B]} = 0 \quad \mbox{or} \quad  R^\ga\,_{\al\ga\be} \bigg[C\Big[e[B]\Big] \bigg]\, 
- \frac{1}{2}\, \eta_{\al\be}\, R^{\ga\de}\,_{\ga\de} \bigg[C\Big[e[B]\Big] \bigg] = 0
\eeq
They can be brought to the usual explicitly diffeomorphism-covariant form
\beq \label{32}
R_{\mu\nu} [g]\, - \frac{1}{2}\, g_{\mu\nu}\, R [g] = 0
\eeq
with the metric
\beq  \label{33} 
g_{\mu\nu} [B]\equiv e^\al\,_\mu [B]\, \eta_{\al\be}\, e^\be\,_\nu [B]
\eeq
defined as usual and $e^\al\,_\mu$ subject to the constraint $x_\al\, e^\al\,_\mu = x_\mu$ which is the only remnant of the fundamental gauge field $\Bagd$.

So in the classical approximation the {\bf SO(1,3)\/} gauge field theory is equivalent to GR and the demonstration that a Schwarzschild solution exists boils down to solving Eqn.(\ref{32}) with $e^\al\,_\mu$ subject to the constraint $x_\al\, e^\al\,_\mu = x_\mu$.

\section{Spherically Symmetric Ansatz for $e^\al\,_\mu$ Fulfilling the Constraint $x_\al\, e^\al\,_\mu = x_\mu$}

\paragraph{}
In this section we specify a general spherically symmetric Ansatz for the inverse Vierbein $e^\al\,_\mu$ compatible with the constraint $x_\al\, e^\al\,_\mu = x_\mu$ -- which in GR does amounts to a coordinate choice.

Let us start with the spherically symmetric time-dependent Ansatz
\beqq  \label{34} 
e^0\,_0 (x^0, r) &=& A (x^0, r) \nonumber \\
e^a\,_0 (x^0, r) &=& B (x^0, r)\, \frac{x^a}{r} \\
e^0\,_m (x^0, r) &=& C (x^0, r)\, \frac{x_m}{r} \nonumber \\
e^a\,_m (x^0, r) &=& D (x^0, r)\, \Bigg( \de^a\,_m\, -\, \frac{x^a\, x_m}{r^2} \Bigg)\, 
+\,E (x^0, r)\, \frac{x^a\, x_m}{r^2} \nonumber
 \eeqq
for the Vierbein $e^\al\,_\mu$ which is inverse to $e_\be\,^\nu$ originally appearing in the theory in Eqn.(\ref{9}). Above $r = x_a\, x^a$ and indices are raised and lowered with $\eta$.

The Ansatz is specified in terms of five undetermined functions $A (x^0, r)$, $B (x^0, r)$, $C (x^0, r)$, $D (x^0, r)$ and $E (x^0, r)$ and is sufficiently general to prove the existence of the Schwarzschild solution. 

Note that for simplicity we have left away a term $F (x^0, r)\, \ep^a\,_{m n}\, \frac{x^n}{r}$ in $e^a\,_m (x^0, r)$ in our Ansatz which is also spherically symmetric and compatible with the constraint.

Imposing the constraint $x_\al\, e^\al\,_\mu (x^0, r) = x_\mu$ on the inverse Vierbein which is easily shown to be equivalent to the one on the original Vierbein Eqn.(\ref{11}) we find the two conditions
\beq
x_\al\, e^\al\,_0 = A\, x_0 + B\, r =\!\!\!\!\!^! \,\,\,\,x_0
\quad \mbox{or} \quad B (x^0, r) = \Big(A (x^0, r) - 1\Big)\, \frac{x^0}{r}
\eeq
and
\beq
x_\al\, e^\al\,_m = C\, \frac{x_0}{r}\, x_m + E\, x_m =\!\!\!\!\!^! \,\,\,\,x_m
\quad \mbox{or} \quad E (x^0, r) = \Big( 1 + C (x^0, r)\, \frac{x^0}{r} \Big)
\eeq
linking $B (x^0, r)$ to $A (x^0, r)$ and $E (x^0, r)$ to $C (x^0, r)$.

As a result the Ansatz Eqns.(34) respecting $x_\al\, e^\al\,_\mu = x_\mu$ becomes
\beqq \label{37}
e^0\,_0 (x^0, r) &=& A (x^0, r) \nonumber \\
e^a\,_0 (x^0, r) &=& \Big(A (x^0, r) - 1\Big)\, \frac{x^0\, x^a}{r^2} \nonumber \\
e^0\,_m (x^0, r) &=& C (x^0, r)\, \frac{x_m}{r} \\
e^a\,_m (x^0, r) &=& D (x^0, r)\, \Bigg( \de^a\,_m\, -\, \frac{x^a\, x_m}{r^2} \Bigg) \nonumber \\
&+& \, \Big( 1 + C (x^0, r)\, \frac{x^0}{r} \Big)\, \frac{x^a\, x_m}{r^2} \nonumber
\eeqq
with $D$ multiplying the transversal projector $\de_a\,^m\, -\, \frac{x_a\, x^m}{r^2}$ and $1 + C\, \frac{x^0}{r}$ the longitudinal one $\frac{x_a\, x^m}{r^2}$.

The crucial point is that the constraint has left us with $A (x^0, r)$, $C (x^0, r)$ and $D (x^0, r)$ undetermined -- sufficient degrees of freedom to demonstrate the existence of the Schwarzschild solution.

\section{The Spherically Symmetric Metric $g_{\mu\nu}$ Corresponding to $e^\al\,_\mu$ and the Line Element $ds^2$}

\paragraph{}
In this section we calculate the metric $g_{\mu\nu}$ in Eqn.(\ref{33}) from $e^\al\,_\mu$ and the corresponding line element $ds^2$.

To prove the existence of a Schwarzschild solution it is easiest to work with the metric which allows us to use standard results in GR. For our Ansatz we find
\beqq \label{38} 
g_{0 0} (x^0, r) &=& -A (x^0, r)^2\, +\, \Big(A (x^0, r)\, -1 \Big)^2\, \bigg(\frac{x^0}{r}\bigg)^2 \nonumber \\
g_{0 m} (x^0, r) &=& g_{m 0} (x^0, r)\,\, =\,\, \Bigg(-A (x^0, r)\, C (x^0, r) \nonumber \\
&+& \Big(A (x^0, r)\, -1 \Big)\, \frac{x^0}{r}\, \bigg(1 + C (x^0, r)\, \frac{x^0}{r} \bigg) \Bigg) \, \frac{x_m}{r} \\
\label{50} g_{m n} (x^0, r) &=& D (x^0, r)^2\, \Bigg( \de_{m n}\, -\, \frac{x_m\, x_n}{r^2} \Bigg) \nonumber \\
&+& \Bigg(-C (x^0, r)^2\, + \bigg( 1 + C (x^0, r)\, \frac{x^0}{r} \bigg)^2 \Bigg)\, \frac{x_m\, x_n}{r^2}. \nonumber
\eeqq

Setting $x^0 = t$ in the sequel we next determine the line element $ds^2$ belonging to the metric $g_{\mu\nu}$ in spherical coordinates
\beqq  \label{39} 
ds^2 &=& -\Bigg( A (t, r)^2\, -\, \Big(A (t, r)\, -1 \Big)^2\, \bigg(\frac{t}{r}\bigg)^2\Bigg) dt^2 \nonumber \\ 
&+&  2 \Bigg(-A (t, r)\, C (t, r)\, \nonumber \\
&+& \Big(A (t, r)\, -1 \Big)\, \frac{t}{r}\, \bigg(1 + C (t, r)\, \frac{t}{r} \bigg) \Bigg)\, dt\, dr \\ 
&+& \Bigg(-C (t, r)^2\, + \bigg( 1 + C (t, r)\, \frac{t}{r} \bigg)^2 \Bigg)\, dr^2 \nonumber \\ 
&+& r^2\, (d\th^2 + \sin^2 \th \,d\varphi^2 ) \nonumber
\eeqq
where we have further set $D (x^0, r) = 1$ to normalize the circumference of a circle to $2 \pi r$.

\section{Existence of the Schwarzschild Solution}

\paragraph{}
In this section we prove the existence of a Schwarzschild solution in the {\bf SO(1,3)\/} gauge field theory of gravitation. We thereby follow the standard approach to demonstrate that a time-dependent spherically symmetric metric Ansatz for the Einstein equations together with Birkhoff's theorem yields the usual static Schwarzschild solution \cite{str}.

We start noting that the line element Eqn.(\ref{39}) is of the general form
\beqq
ds^2 &=& -U(t,r)\, dt^2 + 2\, V(t,r)\, dt\, dr + W(t,r)\, dr^2 \\ 
&+& r^2\, (d\th^2 + \sin^2 \th \,d\varphi^2 ) \nonumber
\eeqq
Defining a new time coordinate
\beq
dt' \equiv \zeta(t,r) \Big(-U(t,r)\, dt + V(t,r)\, dr \Big)
\eeq
we can eliminate the cross-$dt\, dr$ term. Above $\zeta(t,r)$ is an integrating factor the existence of which is demonstrated in the standard literature \cite{str}.

We find
\beqq
ds^2 &=& -\frac{1}{\zeta(t,r)^2\, U(t,r)}\, dt'^2 \nonumber \\
&+& \Bigg(W(t,r) + \frac{V(t,r)^2}{U(t,r)} \Bigg)\, dr^2 \\ 
&+& r^2\, (d\th^2 + \sin^2 \th \,d\varphi^2 ) \quad \mbox{with} \quad t = t(t',r) \nonumber
\eeqq
This can be rewritten as
\beq 
ds^2 = -F(t',r)\, dt'^2 + G(t',r) dr^2 + r^2\, (d\th^2 + \sin^2 \th \,d\varphi^2 )
\eeq

Following the usual steps in the literature inserting the metric into the Einstein equations Eqn.(\ref{32}) they can be solved for $F$ and $G$ which do not depend on time and take the form
\beq
F(t',r) = 1-\, \frac{r_S}{r} = \frac{1}{G(t',r)}
\eeq
where $r_S = 2 \it\Ga\,\! m$ is the Schwarzschild radius with $m$ the mass of the gravitating object.

With a suitable coordinate transformation this standard form of the Schwarzschild line element can be transformed back to the form Eqn.(\ref{38}) and finally Eqn.(\ref{37}) obeying the constraint on $e^\al\,_\mu$ with which we started. This demonstrates the existence of the Schwarzschild solution within the new {\bf SO(1,3)\/} gauge theory which as a result is a proper theory of gravitation at the classical level.

\section{Conclusions}

\paragraph{}
In this paper we have demonstrated the existence of the Schwarzschild solution to the vacuum field equations within the present {\bf SO(1,3)\/} gauge theory in the classical approximation. The proof depends on two crucial facts.

First, the truncated action in the classical limit is a functional of $e[B]$ only. In fact it is the same functional of $e[B]$ as is the Einstein-Hilbert action of the Vierbein $e$ in the respective formulation of GR and there is no explicit dependence on the fundamental field $B$ in the classical approximation. And the resulting field equations are nothing but the vacuum Einstein equations, albeit with the constraint $e_\al\,^\th[B]\, x_\th = x_\al$ on the Vierbein which is the only remnant of it being a composite field $e_\al\,^\th[B] = \eta_\al\,^\th - B_\al\,^{\th\ze}x_\ze$.

Second, the aforementioned constraint allows for a spherically symmetric time-dependent Ansatz with enough degrees of freedom to link it through general coordinate transformations to the standard form used in GR to solve the Einstein equations for the Schwarzschild solution.

We note that the very same approach should work for the determination of other more involved solutions of the truncated {\bf SO(1,3)\/} gauge field theory like cosmological solutions.

As a consequence the {\bf SO(1,3)\/} gauge field theory developed in \cite{chw1} and truncated to the classically dominant term contains the Schwarzschild solution which makes it a perfectly viable theory for gravitation.

And in comparison to GR there are some promising features to the untruncated full theory.

On the classical side perturbative additions to the Schwarzschild solution coming from terms in $S^{(4)}_G$ might account for phenomena usually attributed to "dark matter".

On the quantum field theoretical side the key part is obviously the renormalizability of the full quantum theory proven in \cite{chw3}. Renormalizability in our case -- as opposed to the case of higher derivative theories of gravity expressed in terms of $e$ or $g$ \cite{buc, dono} -- does not come at the price of adding non-physical ghosts which do not decouple from the physical sector of the theory \cite{dono, stel}. Inspite of the absence of such ghosts there remains the unitarity problem related to the non-compact nature of {\bf SO(1,3)\/} \cite{chw2,chw3}.

However, if there is a satisfying answer to the unitarity problem the full quantum theory would bring the description of gravity at par with the description of the electro-weak and strong forces in a unified model based on the principles of gauge invariance and renormalizablity. And it might speculatively even account for the fundamental physical nature of "dark energy" e.g. as some sort of gas of gravitational quanta hyper-weakly interacting with the rest of the universe in some analogy to the cosmic microwave background radiation -- one could say gravity bending over itself as "dark energy".

\end{document}